# Optimizing FIR Filter Mapping on the Morphosys Reconfigurable System

H. Diab and I. Damaj
diab@aub.edu.lb; id01@aub.edu.lb
Department of Electrical and Computer Eng'g
Faculty of Engineering and Architecture
American University of Beirut
Beirut, Lebanon

F. Kurdahi
kurdahi@ece.uci.edu
Department of Electrical and Computer Eng'g
University of California, Irvine
CA 92697
U.S.A.

**ABSTRACT:** This paper proposes an optimized mapping of the FIR filter algorithm that enhances the rate of a reconfigurable computer over a basic mapping previously proposed [1]. It also presents a new interconnection scheme in the reconfigurable part of MorphoSys, a reconfigurable computing system [2]. Reconfigurable computing (RC) is introduced, followed by the MorphoSys RC system. Two optimized FIR mappings are then presented which deliver enhanced speed. A spreadsheet model will detail the modification and the improvement. The speedup achieved is also explained as well as the advantages in the mapping of the application.

## 1. Introduction

Reconfigurable computing (RC) presents an intermediate solution between general-purpose processors (GPPs) and application specific integrated circuits (ASICs). Because it employs hardware that is programmable by software, RC has advantages over both GPPs and ASICs. It provides a faster solution than a GPP because the function that the hardware performs and its interconnection are both specified through software that maps the application onto the reconfigurable hardware.  It has a wider applicability than ASICs since its configuring software makes use of the broad range of functionality supported by the reconfigurable device. It is also cheaper than an ASIC solution since it can use only one piece of hardware to perform the different sub-tasks in a heterogeneous application instead of employing an ASIC for each of those sub-tasks [3]. Achieving higher speeds than GPPs and having more flexibility than ASICs, RC is finding its way more and more into research efforts and hardware devices.

Reconfigurable computing is the next generation for a wide variety of applications such as image processing, image understanding, parallel data computation, encryption, Digital Signal Processing (DSP) and information mining. Particularly RC has been used in DSP, Transforms (e.g. Discrete Cosine Transform) and Filtering (e.g. FIR-Filtering). There has been considerable work for reconfigurable hardware implementations of DSP algorithms and mainly FIR filters as in [7 - 10]. The work done in [7 - 10] concentrates on FPGA implementation of FIR filters with different tap number.

The availability of DSP algorithms supported on reconfigurable coprocessors or programmable chips will open the road to a new generation of reconfigurable DSP processors. The usage of hardware FIR filters, as a part of DSP chips, is important in many fields including mobile communications, communication systems (transceivers), multimedia applications, acoustics, speech, and digital image processing [11 – 13].

## 2. MorphoSys RC System

One of the emerging RC systems includes the MorphoSys designed and implemented at the University of California, Irvine. It has the block diagram shown in Figure 1 [3]. It is composed of 1) an array of reconfigurable cells called the RC array, 2) its configuration data memory called context memory, 3) a control processor (TinyRISC), 4) a data buffer called the frame buffer, and 5) a DMA controller [3].

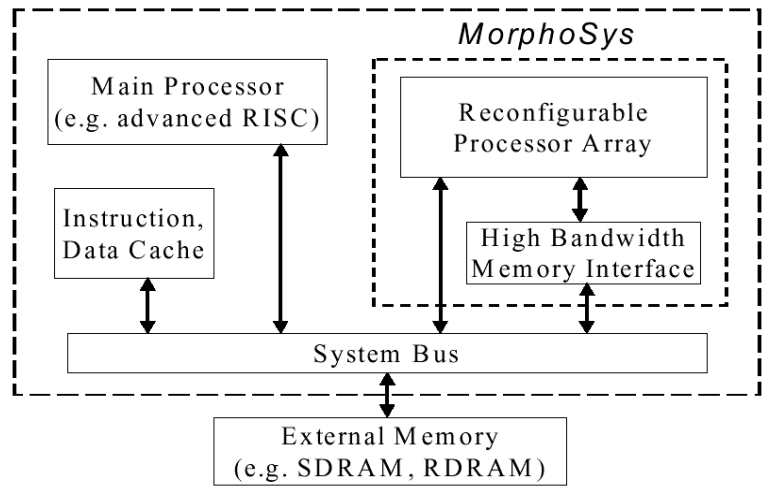

Fig.1. MorphoSys General Architectural Model

A program runs on MorphoSys in the following manner: General-purpose operations are handled by the TinyRISC processor, while operations that have a certain degree of parallelism, regularity, or intensive computations are mapped to the RC array. The TinyRISC processor controls, through the DMA controller, the loading of the context words to context memory. These context words define the function and connectivity of the cells in the RC array. The processor also initiates the loading of application data, such as image frames, from main memory to the frame buffer. This is also done through the DMA controller. Now that both configuration and application data is ready, the TinyRISC processor instructs the RC array to start execution. The RC array performs the needed operation on the application data and writes it back to the frame buffer. The RC array loads new application data from the frame buffer and possibly new configuration data from context memory. Since the frame buffer is divided into two sets, new application data can be loaded into it without interrupting the operation of the RC array. Configuration data is also loaded into context memory without interrupting the RC array operation. This causes MorphoSys to achieve high speeds of execution [3].

The reconfigurable part of MorphoSys is called the RC array. The first implementation of MorphoSys (M1) has an 8x8 RC array [4] as shown in Figure 2 [2]. The computational complexity of the reconfigurable block is introduced in the following features of the RC array. Each of the 64 identical RCs is the basic unit of reconfiguration. The RC array is divided into four quadrants as seen in Figure 2. The interconnection network of the RC array is based on three hierarchical levels. The first is an RC array mesh that connects each RC to its north, south, east and west neighbours. The second level is an intra-quadrant connectivity that connects each RC to all the cells in its row and all the cells in its column as long as it is in the same quadrant. The highest level of interconnection is the express lane (see Figure 3) that allows data transfer from any one cell out of four in a column (or row) of a quadrant to other cells in an adjacent quadrant but in the same column (or row) [5]. The RC array operates in one of two modes: Column context broadcast mode in which all the cells of a column perform the same operation, or row context broadcast mode in which all the cells of a row perform the same operation [6]. Any application to be mapped onto MorphoSys and expected to make use of the reconfigurable core has to use the functions provided by the RCs and specific interconnections that will enhance application performance by increasing the rate of execution of the application. Each RC cell has two ports, A and B, through which it has access to: 1) the output of other cells, 2) the operand data bus, and 3) the internal register file of the RC. The access to the output of the other cells is built on the 3-layer interconnection network presented above [5].

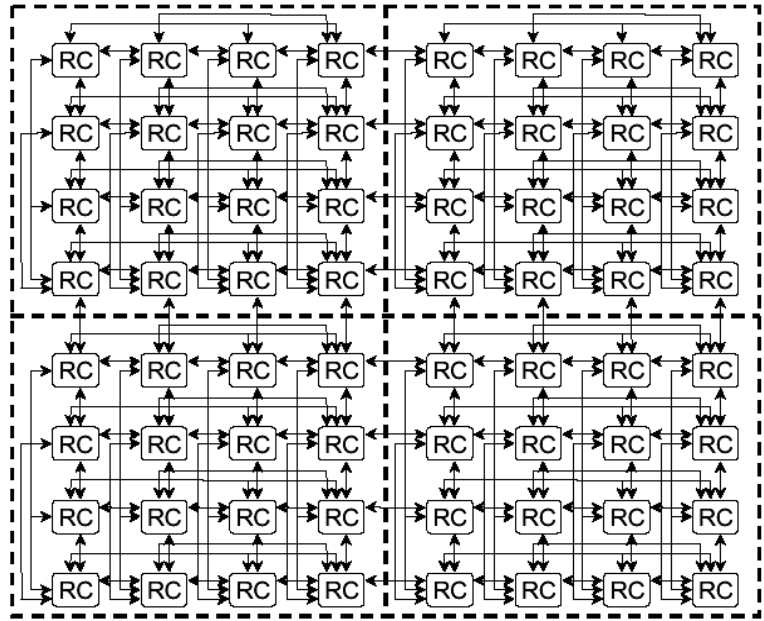

Fig.2. RC-array Interconnection Scheme.

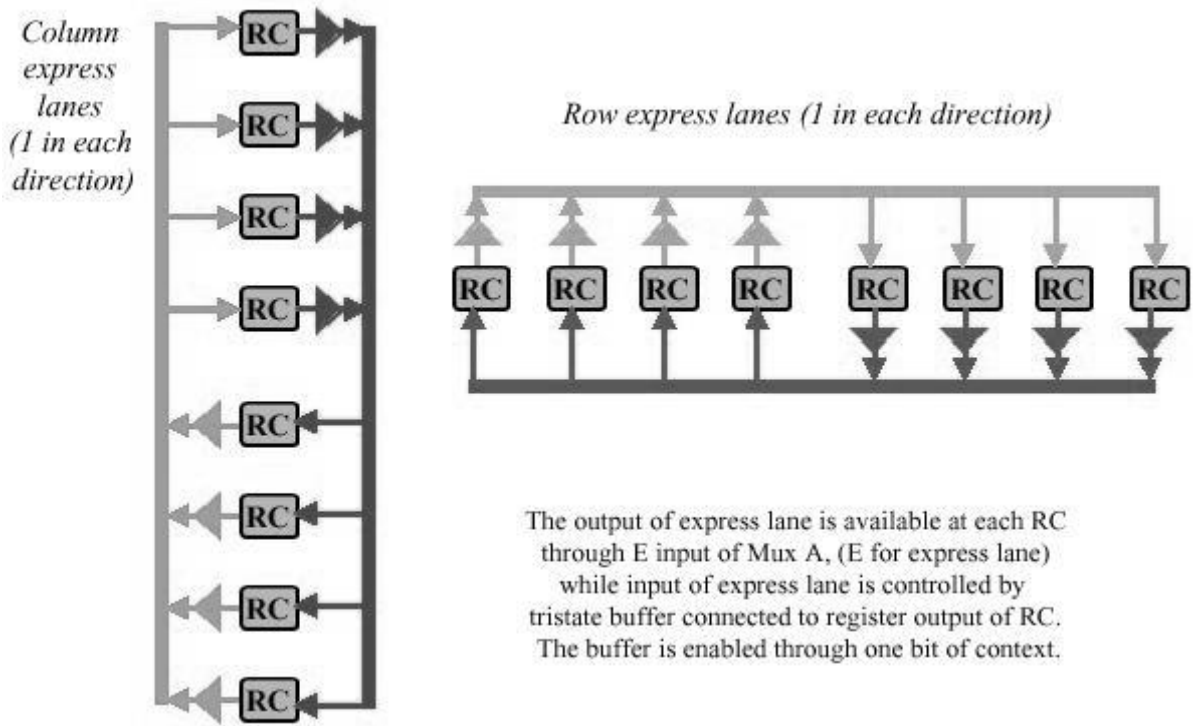

Fig.3. Express lane connectivity

## 3. FIR Filter Basic Mapping

The FIR filter has the following representation: $y_k = \sum_{j=0}^{N-1} x_{k-j} w_j$, or

$$y_k = x_k w_o + x_{k-1} w_1 + x_{k-2} w_2 + ... + x_{k-N-1} w_{N-1}$$

where N is the order of the FIR-filter.

The basic FIR filter mapping suggested in [1] has the model shown in Figure 4. The different context words configuring the columns of the RC array are as follows:

Column 0: Out (t+1) = A x $w_0$ + B, where B accesses the cell to the left.
Column 1: Out (t+1) = A x $w_1$ + B, where B accesses the cell to the left, and
Column 2: Out (t+1) = A x $w_2$,

Where: $w_0$, $w_1$, and $w_2$ are the tap coefficients (or weights). The model presented is for an FIR filter of order 3 and a 3x3 RC array for simplicity and clarity in representation. The idea is general enough though to be applied to an FIR filter of any order and an RC array of any size.

|    | C       | D   | E  | F              | G   | H  | I              | J   | K  | L                    |
|----|---------|-----|----|----------------|-----|----|----------------|-----|----|----------------------|
| 3  | cycle 0 | Data bus | | RC(0,0) = 0 | Data bus | | RC(0,1) = 0 | Data bus | | RC(0,2) = 0 |
| 4  |         |     |    | RC(1,0) = 0    |     |    | RC(1,1) = 0    |     |    | RC(1,2) = 0          |
| 5  |         |     |    | RC(2,0) = 0    |     |    | RC(2,1) = 0    |     |    | RC(2,2) = 0          |
| 6  | cycle 1 | x   | -2 | x-2w2          | x   | -2 | x-2w1          | x   | -2 | x-2w0                |
| 7  |         | x   | -1 | x-1w2          | x   | -1 | x-1w1          | x   | -1 | x-1w0                |
| 8  |         | x   | 0  | x0w2           | x   | 0  | x0w1           | x   | 0  | x0w0                 |
| 9  | cycle 2 | x   | -1 | x-1w2          | x   | -1 | x-1w1+x-2w2    | x   | -1 | x-1w0+x-2w1          |
| 10 |         | x   | 0  | x0w2           | x   | 0  | x0w1+x-1w2     | x   | 0  | x0w0+x-1w1           |
| 11 |         | x   | 1  | x1w2           | x   | 1  | x1w1+x0w2      | x   | 1  | x1w0+x0w1            |
| 12 | cycle 3 | x   | 0  | x0w2           | x   | 0  | x0w1+x-1w2     | x   | 0  | x0w0+x-1w1+x-2w2     |
| 13 |         | x   | 1  | x1w2           | x   | 1  | x1w1+x0w2      | x   | 1  | x1w0+x0w1+x-1w2      |
| 14 |         | x   | 2  | x2w2           | x   | 2  | x2w1+x1w2      | x   | 2  | x2w0+x1w1+x0w2       |
| 15 | cycle 4 | x   | 1  | x1w2           | x   | 1  | x1w1+x0w2      | x   | 1  | x1w0+x0w1+x-1w2      |
| 16 |         | x   | 2  | x2w2           | x   | 2  | x2w1+x1w2      | x   | 2  | x2w0+x1w1+x0w2       |
| 17 |         | x   | 3  | x3w2           | x   | 3  | x3w1+x2w2      | x   | 3  | x3w0+x2w1+x1w2       |
| 18 | cycle 5 | x   | 2  | x2w2           | x   | 2  | x2w1+x1w2      | x   | 2  | x2w0+x1w1+x0w2       |
| 19 |         | x   | 3  | x3w2           | x   | 3  | x3w1+x2w2      | x   | 3  | x3w0+x2w1+x1w2       |
| 20 |         | x   | 4  | x4w2           | x   | 4  | x4w1+x3w2      | x   | 4  | x4w0+x3w1+x2w2       |
| 21 | cycle 6 | x   | 3  | x3w2           | x   | 3  | x3w1+x2w2      | x   | 3  | x3w0+x2w1+x1w2       |
| 22 |         | x   | 4  | x4w2           | x   | 4  | x4w1+x3w2      | x   | 4  | x4w0+x3w1+x2w2       |
| 23 |         | x   | 5  | x5w2           | x   | 5  | x5w1+x4w2      | x   | 5  | x5w0+x4w1+x3w2       |
| 24 | cycle 7 | x   | 4  | x4w2           | x   | 4  | x4w1+x3w2      | x   | 4  | x4w0+x3w1+x2w2       |
| 25 |         | x   | 5  | x5w2           | x   | 5  | x5w1+x4w2      | x   | 5  | x5w0+x4w1+x3w2       |
| 26 |         | x   | 6  | x6w2           | x   | 6  | x6w1+x5w2      | x   | 6  | x6w0+x5w1+x4w2       |
| 27 | cycle 8 | x   | 5  | x5w2           | x   | 5  | x5w1+x4w2      | x   | 5  | x5w0+x4w1+x3w2       |
| 28 |         | x   | 6  | x6w2           | x   | 6  | x6w1+x5w2      | x   | 6  | x6w0+x5w1+x4w2       |
| 29 |         | x   | 7  | x7w2           | x   | 7  | x7w1+x6w2      | x   | 7  | x7w0+x6w1+x5w2       |
| 30 | cycle 9 | x   | 6  | x6w2           | x   | 6  | x6w1+x5w2      | x   | 6  | x6w0+x5w1+x4w2       |
| 31 |         | x   | 7  | x7w2           | x   | 7  | x7w1+x6w2      | x   | 7  | x7w0+x6w1+x5w2       |
| 32 |         | x   | 8  | x8w2           | x   | 8  | x8w1+x7w2      | x   | 8  | x8w0+x7w1+x6w2       |

Fig.4. FIR Filter 3-tap Mapping.

Studying Figure 4, one can notice a considerable amount of redundancy. Cells L14, L16, and L18 all contain the same result. The same is true for cells L17, L19, and L21. Also, for cells: L20, L22, and L24, and so on. Thus, each element in the output vector is actually calculated 3 times (in case of 3-tap filter). In case of the 8-tap filter, each element is calculated 8 times. In fact, there is only a need to calculate each of those values once. Redundancy in Figure 4 can be seen in all columns and not just in the last column. Every 1st and 2nd cells in any column carry the same data as the 2nd and 3rd cells of the same column in the previous cycle. Column 2 in cycle 3 carries the values of $y_0$, $y_1$, and $y_2$, while the same column in cycle 4 carries the values of $y_1$, $y_2$, and $y_3$. This phenomenon is true across all the cycles. Again, this principle is applied to a 3-tap filter for purposes of simplicity, but it is true for any order filter, specifically the 8-tap filter that has been experimented with.

**4. FIR Filter Optimized Mappings**

4.1. FIR Filter Optimized Mapping onto MorphoSys RCs.

To cut back on this redundancy in order to gain some performance, the following adjustment is proposed in the optimized mapping (OM). It keeps the same configuration as the previously suggested mapping, but it requires a rearrangement of the input stream of data. It is interesting to see how keeping the same configuration as before, but slightly rearranging the elements in the input vector, the rate with which the output elements are produced is almost doubled in case of the 3-tap filter and quadrupled in case of the 8-tap filter. This speedup is detailed after the rearrangement is explained.

In Figure 5, the configuration of the cells is the same as in Figure 4. The only change from that figure is the data present on the data bus. With changing input, the contents of the cells in column 2 hold the desired formulas for the output elements. The output elements obtained are shown in column M. The configuration is still the same as that of the basic mapping. In this example, 28 elements in the output vector are obtained in 15 cycles of calculation. The way the input stream needs to be rearranged is as follows:
X = [ $x_0, x_3, x_6, x_1, x_4, x_7, x_2, x_5, x_8, \ldots, x_k, x_{k+3}, x_{k+6}, x_{k+1}, x_{k+4}, x_{k+7}, \ldots$ ]
in the case of the 3-tap filter.

For a higher order filter, the spacing between the elements needs to be the same as the order of the filter. Looking back at Figure 5, it can be seen that cycles 6, 7, 11, and 12 are not producing any results. However, there still is no redundancy here, as these constitute preparatory cycles that calculate intermediate data to be used in later cycles. Some redundancy may be spotted in the cells of column 0. Nothing can be done about this with the given MorphoSys architecture.

|  | C | D | E | F | G | H | I | J | K | L | M |
|---|---|---|---|---|---|---|---|---|---|---|---|
| 3 | cycle 0 | Data bus | | RC(0,0) = 0 | Data bus | | RC(0,1) = 0 | Data bus | | RC(0,2) = 0 | |
| 4 | | | | RC(1,0) = 0 | | | RC(1,1) = 0 | | | RC(1,2) = 0 | |
| 5 | | | | RC(2,0) = 0 | | | RC(2,1) = 0 | | | RC(2,2) = 0 | |
| 6 | cycle 1 | x | 0 | x0w2 | x | 0 | x0w1 | x | 0 | x0w0 | y0 |
| 7 | | x | 3 | x3w2 | x | 3 | x3w1 | x | 3 | x3w0 | |
| 8 | | x | 6 | x6w2 | x | 6 | x6w1 | x | 6 | x6w0 | |
| 9 | cycle 2 | x | 1 | x1w2 | x | 1 | x1w1+x0w2 | x | 1 | x1w0+x0w1 | y1 |
| 10 | | x | 4 | x4w2 | x | 4 | x4w1+x3w2 | x | 4 | x4w0+x3w1 | |
| 11 | | x | 7 | x7w2 | x | 7 | x7w1+x6w2 | x | 7 | x7w0+x6w1 | |
| 12 | cycle 3 | x | 2 | x2w2 | x | 2 | x2w1+x1w2 | x | 2 | x2w0+x1w1+x0w2 | y2 |
| 13 | | x | 5 | x5w2 | x | 5 | x5w1+x4w2 | x | 5 | x5w0+x4w1+x3w2 | y5 |
| 14 | | x | 8 | x8w2 | x | 8 | x8w1+x7w2 | x | 8 | x8w0+x7w1+x6w2 | y8 |
| 15 | cycle 4 | x | 3 | x3w2 | x | 3 | x3w1+x2w2 | x | 3 | x3w0+x2w1+x1w2 | y3 |
| 16 | | x | 6 | x6w2 | x | 6 | x6w1+x5w2 | x | 6 | x6w0+x5w1+x4w2 | y6 |
| 17 | | x | 9 | x9w2 | x | 9 | x9w1+x8w2 | x | 9 | x9w0+x8w1+x7w2 | y9 |
| 18 | cycle 5 | x | 4 | x4w2 | x | 4 | x4w1+x3w2 | x | 4 | x4w0+x3w1+x2w2 | y4 |
| 19 | | x | 7 | x7w2 | x | 7 | x7w1+x6w2 | x | 7 | x7w0+x6w1+x5w2 | y7 |
| 20 | | x | 10 | x10w2 | x | 10 | x10w1+x9w2 | x | 10 | x10w0+x9w1+x8w2 | y10 |
| 21 | cycle 6 | x | 9 | x9w2 | x | 9 | x9w1+x4w2 | x | 9 | x9w0+x4w1+x3w2 | |
| 22 | | x | 12 | x12w2 | x | 12 | x12w1+x7w2 | x | 12 | x12w0+x7w1+x6w2 | |
| 23 | | x | 15 | x15w2 | x | 15 | x15w1+x10w2 | x | 15 | x15w0+x10w1+x9w2 | |
| 24 | cycle 7 | x | 10 | x10w2 | x | 10 | x10w1+x9w2 | x | 10 | x10w0+x9w1+x4w2 | |
| 25 | | x | 13 | x13w2 | x | 13 | x13w1+x12w2 | x | 13 | x13w0+x12w1+x7w2 | |
| 26 | | x | 16 | x16w2 | x | 16 | x16w1+x15w2 | x | 16 | x16w0+x15w1+x10w2 | |
| 27 | cycle 8 | x | 11 | x11w2 | x | 11 | x11w1+x10w2 | x | 11 | x11w0+x10w1+x9w2 | y11 |
| 28 | | x | 14 | x14w2 | x | 14 | x14w1+x13w2 | x | 14 | x14w0+x13w1+x12w2 | y14 |
| 29 | | x | 17 | x17w2 | x | 17 | x17w1+x16w2 | x | 17 | x17w0+x16w1+x15w2 | y17 |
| 30 | cycle 9 | x | 12 | x12w2 | x | 12 | x12w1+x11w2 | x | 12 | x12w0+x11w1+x10w2 | y12 |
| 31 | | x | 15 | x15w2 | x | 15 | x15w1+x14w2 | x | 15 | x15w0+x14w1+x13w2 | y15 |
| 32 | | x | 18 | x18w2 | x | 18 | x18w1+x17w2 | x | 18 | x18w0+x17w1+x16w2 | y18 |
| 33 | cycle 10 | x | 13 | x13w2 | x | 13 | x13w1+x12w2 | x | 13 | x13w0+x12w1+x11w2 | y13 |
| 34 | | x | 16 | x16w2 | x | 16 | x16w1+x15w2 | x | 16 | x16w0+x15w1+x14w2 | y16 |
| 35 | | x | 19 | x19w2 | x | 19 | x19w1+x18w2 | x | 19 | x19w0+x18w1+x17w2 | y19 |
| 36 | cycle 11 | x | 18 | x18w2 | x | 18 | x18w1+x13w2 | x | 18 | x18w0+x13w1+x12w2 | |
| 37 | | x | 21 | x21w2 | x | 21 | x21w1+x16w2 | x | 21 | x21w0+x16w1+x15w2 | |
| 38 | | x | 24 | x24w2 | x | 24 | x24w1+x19w2 | x | 24 | x24w0+x19w1+x18w2 | |
| 39 | cycle 12 | x | 19 | x19w2 | x | 19 | x19w1+x18w2 | x | 19 | x19w0+x18w1+x13w2 | |
| 40 | | x | 22 | x22w2 | x | 22 | x22w1+x21w2 | x | 22 | x22w0+x21w1+x16w2 | |
| 41 | | x | 25 | x25w2 | x | 25 | x25w1+x24w2 | x | 25 | x25w0+x24w1+x19w2 | |
| 42 | cycle 13 | x | 20 | x20w2 | x | 20 | x20w1+x19w2 | x | 20 | x20w0+x19w1+x18w2 | y20 |
| 43 | | x | 23 | x23w2 | x | 23 | x23w1+x22w2 | x | 23 | x23w0+x22w1+x21w2 | y23 |
| 44 | | x | 26 | x26w2 | x | 26 | x26w1+x25w2 | x | 26 | x26w0+x25w1+x24w2 | y26 |
| 45 | cycle 14 | x | 21 | x21w2 | x | 21 | x21w1+x20w2 | x | 21 | x21w0+x20w1+x19w2 | y21 |
| 46 | | x | 24 | x24w2 | x | 24 | x24w1+x23w2 | x | 24 | x24w0+x23w1+x22w2 | y24 |
| 47 | | x | 27 | x27w2 | x | 27 | x27w1+x26w2 | x | 27 | x27w0+x26w1+x25w2 | y27 |
| 48 | cycle 15 | x | 22 | x22w2 | x | 22 | x22w1+x21w2 | x | 22 | x22w0+x21w1+x20w2 | y22 |
| 49 | | x | 25 | x25w2 | x | 25 | x25w1+x24w2 | x | 25 | x25w0+x24w1+x23w2 | y25 |
| 50 | | x | 28 | x28w2 | x | 28 | x28w1+x27w2 | x | 28 | x28w0+x27w1+x26w2 | y28 |

Fig.5. Optimized (3-tap) FIR Filter Mapping.

*4.1.1. Performance Analysis*

For the basic mapping, N results are obtained every N cycles. Thus, the rate of output calculation is 1 per cycle. In the case of the optimized mapping, $N^2$ results are obtained every N+(N-1) cycles (or 2N-1). Thus, a speedup of $N^2/(2N-1)$ is achieved. This speedup versus the filter order is plotted in Figure 6. The basic mapping (BM) of the FIR filter produces N sample outputs every N cycles not counting the write back time. Since write back is made once every N cycles, this mapping produces N sample outputs every (N+1) cycles. Thus, the rate obtained is N/(N+1) as shown in Table 1. The optimized mapping (OM) of the FIR filter produces $N^2$ sample outputs every (2N-1) cycles not counting the write back time. It takes N cycles to write back these results. Thus, the rate obtained is $N^2/(3N-1)$ as shown in Table 1. Table 1 compares the speeds of the basic mapping (BM) and the optimized mapping (OM) for filter orders of 8, 16, 32 and 64. Speeds are in Million Samples per Second (MSPS) also represented as MHz. Table 2 summarizes the above-mentioned rates and speedup for OM and BM.

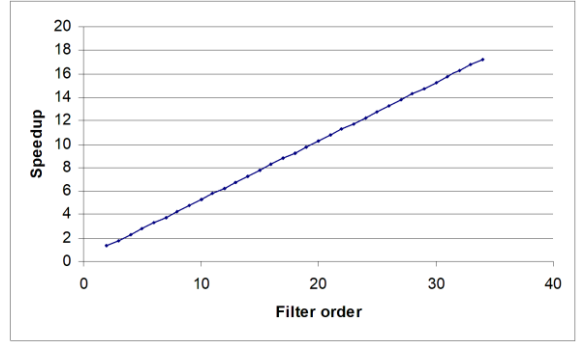

Fig.6. Speedup of Optimized Mapping (OM)

Table 1.
Rates (in MHz) of BM and OM.

|  | Basic Mapping | | | Optimized Mapping | | |
|---|---|---|---|---|---|---|
|  | Samples (N) | Cycles (N+1) | MHz | Samples (N) | Cycles (3N-1) | MHz |
| 8-tap FIR | 8 | 9 | 88.89 | 64 | 23 | 278.3 |
| 16-tap FIR | 16 | 17 | 94.12 | 256 | 47 | 544.7 |
| 32-tap FIR | 32 | 33 | 96.97 | 1024 | 95 | 1077.9 |
| 64-tap FIR | 64 | 65 | 98.46 | 4096 | 191 | 2144.5 |

Table 2.
Rates for BM and OM and Speedup of OM over BM.

|  | Basic Mapping Rate | Optimized Mapping Rate | Speedup of OM over BM |
|---|---|---|---|
| Without considering write back | N/N | $N^2/(2N-1)$ | $N^2/(2N-1)$ |
| With considering write back | N/(N+1) | $N^2/(3N-1)$ | $N(N+1)/(3N-1)$ |

Table 3 reflects the speedup factor ($S_f$) achieved by employing the optimized mapping over the basic mapping for filter orders 8, 16, 32 and 64 taking write back overhead into consideration. These values can be easily computed from the results in Table 1 or the speedup value shaded in Table 2. For example, the speedup value for an 8-tap FIR filter could be calculated as the ratio of (278.3 MHz/ 88.89 MHz), or by substituting 8 for N in the equation N(N+1)/(3N-1).

Table 3.
Speedup of OM over BM.

|  | ($S_f$) Speedup Factor |
|---|---|
| 8-tap FIR | 3.13 |
| 16-tap FIR | 5.79 |
| 32-tap FIR | 11.12 |
| 64-tap FIR | 21.78 |

*4.1.2. Performance Improvement*

The proposed optimized mapping thus produces two results every cycle compared to the basic mapping in [1] that achieved a rate of one result per cycle. This obviously doubles the rate of producing output elements at the last column.

Assuming that the RC has only two input ports, then extending the connection to include newer cells would not enhance the performance any further because those new interconnections would not be made use of. With an extended interconnection, further enhancement on the rate can be achieved if each cell is provided with three input ports and an ALU-multiplier adder of also three inputs that can perform a multiplication of one input and addition with the two other inputs.

4.2. FIR Filter Improved Mapping onto Enhanced MorphoSys Interconnection

A proposed addition to the interconnection network is a simple one. Its goal is to enhance the performance of the FIR filter operation on MorphoSys. The basic FIR filter mapping in [1] has the RCs using port A to access the data bus and port B to access the output of the cell to the left of the specific RC.

Port B accesses the output of the cell to its top, bottom, or left. In this paper, we suggest the addition of a connection that allows port B to access the cell at its lower left corner (diagonally). This is shown in Figure 7 for only one sample cell. This will allow data to travel from a certain cell to a diagonal cell on its upper right corner. This shall prove to enhance performance for the FIR application as shown in section 4.2.2.

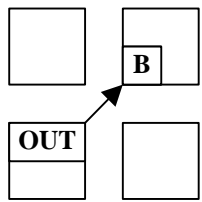

Fig. 7. Suggested Interconnection Addition.

A new mapping, improved mapping (IM), is introduced that will take advantage of this new interconnection. The constraint here is that the maximum order that the filter can assume needs to be less than the RC length or width. In an 8x8 RC array MorphoSys configuration, for example, an FIR filter with a maximum order of 7 can be mapped. In a 16x16 RC array configuration, a maximum order of 15 can be mapped and so on. This constraint will become clearer as the mapping is explained. The new model is explained via spreadsheets for a 3-tap filter in a 4x4 RC array MorphoSys for simplicity and clarity. The same concept may, however, be applied to a filter of any order and an RC array of any size. Figure 8 shows the suggested mapping.

| | C | D | E | F | G | H | I | J | K | L | M | N | O | P | S | V | Y | Z |
|---|---|---|---|---|---|---|---|---|---|---|---|---|---|---|---|---|---|---|
| 3 | cycle 0 | | | | | | RC(0,0) = 0 | | | RC(0,1) = 0 | | | RC(0,2) = 0 | | RC(0,3) = 0 | w2 | w1 | w0 | | |
| 4 | | Data | | | RC(1,0) = 0 | Data | | RC(1,1) = 0 | Data | | RC(1,2) = 0 | Data | | RC(1,3) = 0 | w2 | w1 | w0 | | |
| 5 | | bus | | | RC(2,0) = 0 | bus | | RC(2,1) = 0 | bus | | RC(2,2) = 0 | bus | | RC(2,3) = 0 | w2 | w1 | w0 | | |
| 6 | | | | | RC(3,0) = 0 | | | RC(3,1) = 0 | | | RC(3,2) = 0 | | | RC(3,3) = 0 | w2 | w1 | w0 | | |
| 7 | cycle 1 | | x | -2 | x-2w2 | x | -2 | x-2w1 | x | -2 | x-2w0 | x | -2 | | w2 | w1 | w0 | | |
| 8 | | | x | -1 | x-1w2 | x | -1 | x-1w1 | x | -1 | x-1w0 | x | -1 | | w2 | w1 | w0 | | |
| 9 | | | x | 0 | x0w2 | x | 0 | x0w1 | x | 0 | x0w0 | x | 0 | | w2 | w1 | w0 | | |
| 10 | | | x | 1 | x1w2 | x | 1 | x1w1 | x | 1 | x1w0 | x | 1 | | w2 | w1 | w0 | | |
| 11 | cycle 2 | | x | 0 | x0w2 | x | 0 | x0w1+x-1w2 | x | 0 | x0w0+x-1w1 | x | 0 | | w2 | w1 | w0 | y0 | |
| 12 | | | x | 1 | x1w2 | x | 1 | x1w1+x0w2 | x | 1 | x1w0+x0w1 | x | 1 | | w2 | w1 | w0 | y1 | |
| 13 | | | x | 2 | x2w2 | x | 2 | x2w1+x1w2 | x | 2 | x2w0+x1w1 | x | 2 | | w2 | w1 | w0 | | |
| 14 | | | x | 3 | x3w2 | x | 3 | x2w1 | x | 3 | x3w0 | x | 3 | | w2 | w1 | w0 | | |
| 15 | cycle 3 | | x | 2 | x2w2 | x | 2 | x1w1+x0w2 | x | 2 | x2w0+x1w1+x0w2 | x | 2 | | w2 | w1 | w0 | y2 | |
| 16 | | | x | 3 | x3w2 | x | 3 | x2w1+x1w2 | x | 3 | x3w0+x2w1+x1w2 | x | 3 | | w2 | w1 | w0 | y3 | |
| 17 | | | x | 4 | x4w2 | x | 4 | x4w1+x3w2 | x | 4 | x4w0+x2w1 | x | 4 | | w2 | w1 | w0 | | |
| 18 | | | x | 5 | x5w2 | x | 5 | x5w1 | x | 5 | x5w0 | x | 5 | | w2 | w1 | w0 | | |
| 19 | cycle 4 | | x | 4 | x4w2 | x | 4 | x4w1+x3w2 | x | 4 | x4w0+x2w1+x1w2 | x | 4 | | w2 | w1 | w0 | y4 | |
| 20 | | | x | 5 | x5w2 | x | 5 | x5w1+x4w2 | x | 5 | x5w0+x4w1+x3w2 | x | 5 | | w2 | w1 | w0 | y5 | |
| 21 | | | x | 6 | x6w2 | x | 6 | x6w1+x5w2 | x | 6 | x6w0+x5w1 | x | 6 | | w2 | w1 | w0 | | |
| 22 | | | x | 7 | x7w2 | x | 7 | x7w1 | x | 7 | x7w0 | x | 7 | | w2 | w1 | w0 | | |
| 23 | cycle 5 | | x | 6 | x6w2 | x | 6 | x6w1+x5w2 | x | 6 | x6w0+x5w1+x4w2 | x | 6 | | w2 | w1 | w0 | y6 | |
| 24 | | | x | 7 | x7w2 | x | 7 | x7w1+x6w2 | x | 7 | x7w0+x6w1+x5w2 | x | 7 | | w2 | w1 | w0 | y7 | |
| 25 | | | x | 8 | x8w2 | x | 8 | x8w1+x7w2 | x | 8 | x8w0+x7w1 | x | 8 | | w2 | w1 | w0 | | |
| 26 | | | x | 9 | x9w2 | x | 9 | x9w1 | x | 9 | x9w0 | x | 9 | | w2 | w1 | w0 | | |

Fig.8. Mapping of the Improved Design.

In this mapping, the last column is not made use of. So, a 4x3 RC array configuration could work, but since the RC array is usually of equal length and width, then a 4x4 RC array is considered in this specific case and an NxN RC array in general for a filter of maximum order (N-1). Column 0 (spreadsheet column F), as in the case of the mapping in [1], multiplies the input by weight $w_2$. Columns 1 and 2 (spreadsheet columns I and L in Figure 8) multiply the input by weights $w_1$ and $w_2$ respectively and add the result to the output of the cell on its lower left corner. This is where the suggested new interconnection is made use of. The last column, column 3 in this case, is not needed. Note that the cells in the last row of columns 1 and 2 do not have a cell to their lower left corner, so the result of the addition is added to zero. Notice also that the order of the input elements is sequential with no need for rearrangement.

Looking at column L in the model, i.e. column 2, one can see that the first two cells in each cycle hold the formulas for two output vector elements. The specific elements can be seen in column Z. This diagonal communication from a cell to the cell on its upper right corner can be seen clearly throughout the figure.

*4.2.1. Performance Improvement*

This mapping thus produces two results every cycle compared to the basic mapping in [1] that achieved a rate of one result per cycle. This obviously doubles the rate of producing output elements at the last column. Assuming that the RC has only two input ports, then extending the connection to include newer cells would not enhance the performance any further because those new interconnections would not be made use of. With an extended interconnection, further enhancement on the rate can be achieved if each cell is provided with three input ports and an ALU multiplier/adder of also three inputs that can perform a multiplication of one input and addition with the two other inputs.

*4.2.2. Performance Analysis*

For the basic mapping of the FIR filter in [1], N results are obtained every N cycles. Thus, the rate of output calculation is 1 per cycle. In the case of the improved mapping, 2 results are obtained every cycle giving a rate of 2 samples per cycle. Thus, a speedup of 2 is achieved in the improved design over the basic design. This speedup is in the production of output samples in the contents of the RCs of the last column in the RC array. It does not take into consideration the time taken to write back the results to the frame buffer.

For the basic FIR filter mapping in [1], write back is made once every N cycles, thus producing N sample outputs every (N+1) cycles. The rate obtained is N/(N+1). For the FIR filter mapping on the improved design, write back is made once every cycle, thus producing 2 output samples every 2 cycles with a rate of 1 sample per cycle.

Table 4 shows the number of samples obtained for a specific number of cycles for both the FIR filter basic mapping (BM) in [1] and the improved mapping (IM) design proposed above. Table 4 also includes the rates achieved for the FIR filter of orders 8, 16, 32 and 64. The rate is in Million Samples per Cycle (MSPS) or MHz assuming that MorphoSys is running at a clock frequency of 100 MHz. The FIR filter mapping using the IM design achieves a fixed rate of 100 MHz for an FIR filter of any order. Thus, it has better performance than the basic FIR filter mapping achieving speeds between 88 and 98 MHz for filter orders between 8 and 64. Table 5 summarizes the above-mentioned rates for BM and IM. Table 6 shows the speedups; these values can be easily computed from the results in Table 4 or the speedup value shaded in Table 5. For example, the speedup value for an 8-tap FIR could be calculated as the ratio of (100 MHz / 88.89 MHz), or by substituting 8 for N in the shaded cell in Table 5.

Table 4.
Rates (in MHz) of BM and IM

|  | Basic Mapping | | | Improved Mapping | | |
| --- | --- | --- | --- | --- | --- | --- |
|  | Samples (N) | Cycles (N+1) | MHz | Samples (N) | Cycles (N) | MHz |
| 8-tap FIR | 8 | 9 | 88.89 | 2 | 2 | 100 |
| 16-tap FIR | 16 | 17 | 94.12 | 2 | 2 | 100 |
| 32-tap FIR | 32 | 33 | 96.97 | 2 | 2 | 100 |
| 64-tap FIR | 64 | 65 | 98.46 | 2 | 2 | 100 |

Table 5.
Rates for BM and IM and Speedup of IM over BM.

|  | Basic Mapping Rate | Improved Mapping Rate | Speedup of /IM over BM |
|---|---|---|---|
| Without considering write back | N/N = 1 | 2N/N = 2 | 2 |
| With considering write back | N/(N + 1) | 2N/2N = 1 | (N + 1)/N |

Table 6.
Speedup of IM over BM.

|  | ($S_f$) Speedup Factor |
|---|---|
| 8-tap FIR | 1.24 |
| 16-tap FIR | 1.062 |
| 32-tap FIR | 1.031 |
| 64-tap FIR | 1.015 |

4.3. Performance Analysis: Comparisons with Similar Work

With the emergence of powerful reconfigurable systems, FIR filters implementations where carried in different research groups [7 – 10] and industrial firms (e.g. Xilinx). Some of that work was done in the University of Kansas for synthesizing an efficient FIR filter reconfigurable architecture using FPGAs. The FPGA implementation suggests that 60-70 tap chips with sampling rates exceeding 100 MHz should be feasible [9].

Another automatic implementation of FIR filters on FPGAs is given in [10]. The presented structure gives the best performance with a maximum sampling rate of 33.3 MHz, 31.8 MHz and 31.3 MHz for an 11-tap FIR filter. For a 19-tap linear phase filter as in [9] the sampling rates were in the order of 15-20 MHz. Table 7 shows these figures compared with the MorphoSys improved FIR-filter mapping..

Table 7.
A comparison between different reconfigurable hardware implementations of FIR-filters.

| Filter Type | MorphoSys (Improved Mapping) | FPGAs (Xilinx) | Speedup |
|---|---|---|---|
| **11-tap FIR filter** | 100 MHz | 30 MHz (XC3195) | 3.3 |
| **11-tap FIR filter** | 100 MHz | 33.3 MHz (XC3195) | 3 |
| **19-tap FIR filter** | 100 MHz | 15-20 MHz (XC4020) | 5 |

As can be seen from Table 7, the estimated speedup of MorphoSys over the XC3195 FPGA [9-10] is over 3 for the 11-tap FIR filter case, and at least 5 for the 19-tap FIR filter when compared with the performance of the XC4020 [9] FPGA. It is worth mentioning here that while the MorphoSys system clock is 100MHz, that of the XC3195 and XC4020 FPGAs are 85 and 80 MHz respectively.

**5. Conclusion**

Many advantages could be established from the proposed findings. Firstly, the implementation of FIR filters using dynamically reconfigurable hardware. Secondly, the realization of an optimized mapping of FIR filters on MorphoSys. Thirdly, results out of the optimized mapping are advantageously compared to the basic mapping with respect to speed up, and the rate of output sample production. Fourthly, findings are compared with another reconfigurable hardware implementation, namely the Xilinx FPGAs. Moreover, high speedup factors where achieved by the optimized mapping over the basic mapping. The key to this speedup is an increased utilization of the parallelism offered by both MorphoSys and the FIR filter computation. This was done by rearranging the input stream of elements pertaining to the FIR filter. The output elements are produced in the same order as the input stream. Rearrangement of input and output elements is needed in case the input elements are provided in the correct order and the output desired in the correct order as well. An improved mapping is also proposed which further enhances the system performance for FIR implementation, which required an interconnection upgrade to the MorphoSys system.

H. Diab, **I. Damaj**, F. Kurdahi, Optimizing FIR Filter Mapping on MorphoSys, *The International Journal of Parallel and Distributed Systems and Networks,* ACTA press, 2002. I 3, V 5, P 108 – 115.